\documentstyle[aps,prl,epsf]{revtex}
\begin{document}
\draft
\twocolumn[\hsize\textwidth\columnwidth\hsize\csname @twocolumnfalse\endcsname
\title{Two-dimensional randomly-frustrated spin 1/2 Heisenberg model}
\author{J. Oitmaa and O.P. Sushkov}
\address{School of Physics, University of New South Wales, Sydney 2052, 
Australia}

\maketitle

\begin{abstract}
We investigate the properties of $S=1/2$ Heisenberg clusters with random
frustration using exact diagonalizations. 
This is a model for a quantum spin glass. We show that the average ground 
state spin is  $S \propto \sqrt{N}$, where $N$ is the number of sites. 
We also calculate the magnetic susceptibility and the spin stiffness
and low-energy excitations, and discuss these in terms of a semiclassical
picture.
\end{abstract}

]

Frustrated quantum spin systems continue to attract much interest. Such
systems show rich and complex behavior, including complex ground
states, with or without long-range magnetic order, quantum phase
transitions at T=0, and novel excitations/quasiparticles. Most work has
focussed on model systems with translation symmetry, in which the
frustration is geometric in origin, resulting from competing
antiferromagnetic exchange interactions. Typical examples are the
triangular lattice \cite{Az,Bern} and the $J_1-J_2$ model on the square lattice
\cite{Dag,Capr,Kotov}. We refer to such systems as "regularly frustrated". 
A recent review of phase transitions in frustrated antiferromagnets 
\cite{Kaw} discusses many of these issues, particularly for 3-d systems. 
A general field theoretic approach is discussed in recent book \cite{Sachdev}.

In contrast to these cases are systems where defects and/or random
interactions or fields lead to frustration. Doty and Fisher \cite{Dot}
investigated the effects of both random fields and random exchange in
the spin-1/2 XXZ chain, and found a quantum phase transition controlled
by exchange anisotropy. Sandvik \cite{Sandvik} considered a square-lattice
antiferromanget with up to 10\% ferromagnetic bonds, but was unable to
reach the interesting spin-glass regime. There have also been a number
of studies of randomly diluted systems, but since these are not, in
general, frustrated we do not discuss them further.

The most extreme case of frustration by exchange disorder occurs in
spin-glass systems \cite{Bind}. There is an enormous body of theoretical work
on models of spin-glasses, but the bulk of this is on Ising or classical
XY and Heisenberg models. The earliest work on quantum spin-glasses, an
attempt to apply replica theory to the spin-S quantum model with
long-ranged Gaussian interactions, is due to Bray and Moore \cite{Bray}. The
same model has been considered more recently via an SU(N) generalization
\cite{Ye}, but the relationship of these studies to the more realistic
short-range interaction case remains unclear. To our knowledge the only
study of a finite S short-range spin-glass model is the work of Nonomura
and Ozeki \cite{Non}, who carried out exact diagonalizations on small clusters.

The present Letter is strongly motivated by Ref.\cite{Non}, and extends that
work in several ways. We consider small clusters of S=1/2 spins with
Hamiltonian
\begin{equation}
\label{H}
H=\sum_{<ij>}J_{ij}{\bf S}_i\cdot{\bf S}_j,
\end{equation}
where $J_{ij}$ are random exchange interactions. Two models for
randomness are used: the $\pm J$ model with
\begin{equation}
\label{P1}
P(J)={1\over{2}}\delta(J-1)+{1\over{2}}\delta(J+1),
\end{equation}
and the Gaussian model with
\begin{equation}
\label{P2}
P(J)={1\over{\sqrt{2\pi}}}e^{-J^2/2}.
\end{equation}
with mean 0 and standard deviation 1. We use exact diagonalizations for
clusters of N=10,16,18,20 spins with periodic boundary conditions. These
clusters have the symmetry of the square lattice, and hence the results
can be extrapolated, using finite size scaling, to the infinite lattice.
We have also carried out some calculations with free boundaries, in
which case our results may be relevant to magnetic nanoparticles.

The first question to be addressed is the average spin of the ground
state. Using at least 500 different randomly selected bond
configurations for each cluster size, we determine the ground state spin
quantum number S for each realization, and the mean value and standard
deviation. Of course, for the homogeneous antiferromagnet S=0
rigorously, but for random interactions we find S values from 0 to 6
(for the larger clusters). Figure 1 shows histograms for N=18 (periodic
boundary conditions and 2000 samples) for both the $\pm J$ and Gaussian
cases, together with the distribution which would be expected in the
absence of interactions
(the number of combinations by which one can add 18 spins 1/2 to get the
total spin $S$).
The histogram for the Gaussian model is slightly shifted to larger $S$
compared to that for the $\pm J$ model, and the pure statistical distribution
is shifted further up. However, all in all the distributions are very similar.

From the statistics of ground state spin quantum numbers we have
computed the average value of ${S^2}$. One can say that this is an
average value of the maximum z-projection of the spin squared. This
quantity has a more straightforward semiclassical meaning
than ${\overline {<S^2>}}={\overline {S(S+1)}}$. In the
thermodynamic limit, $N \to \infty$ these two quantities coincide.
The value of ${\overline {S^2}}$ is plotted in Fig.2 versus N, the number 
of spins in the cluster. While there is some scatter, due both to limited
statistics and to the different orientations of clusters, the data is
consistent with a linear dependence of ${\overline {S^2}}$ on N
shown by the dashed lines. In particular
\begin{eqnarray}
\label{Sa}
\pm J \ model:&& \ \ \ {\overline {S^2}} \approx 0.120 N,\\
Gaussian \ model:&& \ \ \ {\overline {S^2}} \approx 0.202 N.\nonumber
\end{eqnarray}
This means that the average spin per site varies as $ 1/\sqrt{N}$, 
and vanishes in the thermodynamic limit. This N dependence is exactly what
would be expected from classical fluctuation theory, for a system of
spins with random orientations, though our system is in the extreme
quantum limit. If we denote the length of the effective classical spin by
$S_{eff}$ then $<S^2>=N \ S_{eff}^2$. This yields $S_{eff}$=0.35 for the 
$\pm J$ case and 0.45 for the Gaussian case, compared to  1/2 for
noninteracting classical spins. The quantity is reduced due to quantum 
fluctuations, and we observe that quantum fluctuations are greater in the 
$\pm J$ model than in the Gaussian case.

From the same calculations we obtain the ground state energy per spin
$E_0/N$, averaged over at least 500 independent configurations. These 
values are given in Table 1. There is little variation between different 
clusters, indicating that these results are close to the thermodynamic 
limit. As expected, the Gaussian model yields slightly higher energies.

We have also computed the average spin-glass order parameter in the
ground state
$ m^2_{sg}={1\over{N^2}}{\overline {\sum_{i,j}<{\bf S_i}\cdot {\bf S_j}>^2}},$
where the line denotes a configurational average. The values
of this quantity, averaged over 500 random bond configurations, are
given in Table 1. Plotting the results versus $1/N$ shows a linear
variation, which, when extrapolated to the bulk limit, yields $m_{sg}^2$ =
0.01 for both $\pm J$ and Gaussian cases. This value agrees with the
estimate in Ref. \cite{Non}. The effective spin length $S_{eff}$ can be 
estimated from the spin-glass order parameter, via
$m^2_{sg}=S_{eff}^4<cos^2 \theta>=S_{eff}^4/3$,
which gives $S_{eff} = 0.42$, consistent with the estimate above.

As well as the ground state we have studied properties of the low-lying
excitations of the finite clusters. We have determined
numerically the spin quantum number of the lowest excited state and the
energy gap for at least 500 realizations of bond configurations for
N=10,16,18,20 as before. We observe that predominantly (typically 97\%
of cases) the spin of the first excitation is $S_1 = S_0 \pm 1$ where $S_0$ 
is the spin of the ground  state. In Table 1 we present $f_{+}$ and $f_{-}$: 
the fraction of configurations which have $S_1=S_0+1$ and which have 
$S_1=S_0-1$, for both $\pm J$ and Gaussian models. 
We also give the average energy gap ${\overline {\Delta E}}$
for both models, for those cases where  $S_1=S_0\pm 1$.
The average gap for $S_1=S_0+1$ is practically the same as that
for $S_1=S_0-1$. The frequency $f_-$ appears systematically less than 
$f_+$. We give an explanation of this below. 

The results for the lowest excitations can be understood within a 
semi-classical picture, valid for large N.  We denote by $E_0$ 
and $|0>$ the energy of the ground state, which has total spin
$S_0\propto \sqrt{N}$, and the corresponding wave function with maximum
$S_z=S_0$. We now consider an external
magnetic field B in the z-direction. The system will adopt a new ground 
state, with energy
\begin{equation}
\label{EB}
E_B=E_0-S_0B-{1\over{2}}\chi NB^2,
\end{equation}
where $\chi$  is the magnetic susceptibility per site.
The state $|0>$, which will precess about the z-axis with
Larmor frequency $\omega=B$, is a ``rigid body'' excitation in which
the spins precess coherently. Hence its energy with respect to $E_B$ 
is simply  $\delta E = S_0B + {1\over{2}}I\omega^2$. 
Comparing this with (\ref{EB}) gives the effective moment of
inertia $I=N\chi$. Similar considerations can be also applied
to the 2D quantum antiferromagnet and lead to the known
results \cite{Neub}.

The lowest excitations thus correspond to the states of a 2D quantum rotor
with energies
\begin{equation}
\label{rig1}
E_L-E_0={{L^2}\over{2I}}={{L^2}\over{2N\chi}},
\end{equation}
where $L=0,\pm 1, \pm 2,...$ is the angular momentum. The total spin of the
corresponding excited state is $S=S_0+L$.
Thus the lowest excitations will have $S_1=S_0\pm 1$, as observed
in the numerical results.
Note that eq. (\ref{rig1}) describes rotations around the z-axis. 
This is why it is different from that for the antiferromagnet \cite{Neub}
which describes 3D rotations.
In addition to the requirement $N \gg 1$ for this approach, we also require
$|L| \ll S_0 \sim \sqrt{N}$ to  justify the formula (\ref{rig1}).
In our numerical calculations this inequality is only weakly
satisfied, and we believe this is the reason for the number of samples
 with $S_1=S_0-1$ being somewhat less than with $S_1=S_0+1$.

In the above consideration the susceptibility $\chi$ corresponds to the
magnetic field directed along the total cluster spin. We argue that the
susceptibility determined in this way from the cluster data extrapolates,
in the thermodynamic limit, to the total susceptibility and not to
$\chi_{\perp}$ (as in the antiferromagnet). This is because, in the
$N\to \infty$ limit, the $B^2$ term in (\ref{EB}) will be dominant, and
the energy will be independent of the spin direction.

 From (\ref{rig1}) we also see that the energy gap should behave like $1/N$. 
In Fig.3   we plot the quantity $1/2\chi = N \ {\overline {\Delta E}}$
versus $1/N$. Although finite size corrections appear to be large, a rough
extrapolation to $N=\infty$  gives for the magnetic susceptibility
$\chi\approx 0.40$ ($\pm J$ model), and $\chi\approx 0.91$ (Gaussian). 
These values are about an order of magnitude higher than for the 
2D quantum S=1/2 antiferromagnet. 
Another interesting point is that the finite size correction is of the
opposite sign compared to that for the 2D quantum  antiferromagnet.

The final quantity we have computed is the spin-stiffness  $\rho_s$, which
is a measure of the rigidity of the ground state to a small twist $\theta$,
\begin{equation}
\label{rho}
\delta E_0 = {1\over{2}}\rho_s \int \left(\nabla \theta\right)^2 d^2r,
\end{equation}
where $E_0$ is the ground state energy.
Non-zero spin-stiffness, in the thermodynamic limit,
indicates a long-range magnetic order in the system.
Previous calculations of $\rho_s$ for finite antiferromagnetic
clusters \cite{Bonka} have used a modified Hamiltonian in which the
quantization axis in the twist direction is rotated by  $\Delta\theta$ for
neighboring sites, leading to phase factors in the exchange constants.
This method of calculation assumes that the system has an intrinsic spatial
periodicity, so it is applicable to ferromagnetic or antiferromagnetic
states. However we found that it cannot be applied to a  random 
system. This is why we use a different approach.
As the exact diagonalization method yields not only the ground state 
energy but also the ground state wavefunction $\Psi$ for each cluster, we 
impose the twist directly on the wave function, 
$\Psi_{\theta}=U_{\theta} \Psi$, where $U_{\theta}$ is the unitary 
transformation which rotates each subsequent spin along a given direction
by an additional angle $\Delta{\theta}$. The change of the energy under the 
twist,  $\delta E_0 = <\Psi_{\theta}|H|\Psi_{\theta}>-E_0$ is calculated
explicitly, and comparison with (\ref{rho}) gives the spin stiffness.
To avoid ambiguity we have chosen clusters with free boundaries, and with
the twist imposed along the $x$  direction. Table 2 presents
values of $\rho_s$ for various clusters, averaged over 500 bond
configurations. There is a weak decrease of $\rho_s$ with size of the 
cluster, but the scatter due to different shapes does not allow us to make a 
reliable extrapolation.
Therefore we take an average over clusters value as an estimate 
for the stiffness.  We have also performed similar calculations for 
antiferromagnetic clusters and find that the data is very similar: 
a small scattering and a weak decrease with size. However in this case
we know the thermodynamic value and hence find that our cluster
calculations overestimate the stiffness by the factor 1.4. Assuming that
the finite size scaling factor for the random system is the same
we come to the following estimates for the stiffness, 
$\rho_s \approx 0.15$ ($\pm J$ model), and $\rho_s \approx 0.14$ 
(Gaussian model).    
These values are just slightly less than the spin stiffness for the
2D quantum antiferromagnet on the square lattice.

Having the spin stiffness and the magnetic susceptibility we find
the velocity of the Goldstone spin wave $c=\sqrt{\rho_s/\chi}$, (see
Ref. \cite{Halperin,Bind}), $c\approx 0.6$ ($\pm J$ model) and 
$c\approx 0.4$ (Gaussian model). This should be compared with
$c=1.67$ for the 2D quantum antiferromagnet on the square lattice.
The low temperature specific heat will be due to the Goldstone spin
waves and thus will have a $T^2$ temperature dependence. We propose
that this will be generic for disordered isotropic 2D quantum spin
systems, and explains, for example, the data for the $S=3/2$ Kagome 
spin glass \cite{Ramirez}. This should be contrasted with Ising spin
glasses where the specific heat is due to the localized excitations,
and is linear in $T$.

In summary, we have investigated aspects of highly disordered S=1/2
quantum spin systems at zero temperature by means of exact
diagonalizations of small clusters of up to 20 spins. Averaging over
disorder, we find: 1)the total spin quantum number scales as 
$S\propto \sqrt{N}$, 2)there is evidence of spin-glass order in the ground 
state, in the thermodynamic limit,
3)the lowest excitations and energy gap for clusters are consistent with the
predictions of a semi-classical theory,
4)the spin-stiffness does not appear to scale to zero in the
      thermodynamic limit.
Two models for the random exchange have been used, a $\pm J$ model and
one with Gaussian exchange. There appears to be no significant
qualitative difference between these two cases. Although we have used
square lattice clusters with nearest-neighbor interactions and with
periodic boundary conditions, to model the behavior of an infinite
square lattice, we believe our results are more general.

There are two ways in which our results could be related to real
systems. The first is to quantum spin-glasses, which have received
little attention in the literature. However, one needs to be cautious in
extrapolating from small clusters to systems as complex as spin-glasses.
An alternative application might be to disordered magnetic
nanoparticles. We know of no immediate candidates. However recent work
on  Mn$_{12}$ and Fe$_{13}$O$_3$ spin clusters \cite{Kat} 
(which are not disordered) suggests that this is not impossible.

Finally we remark on a recent study of a system of N randomly
interacting fermions \cite{Mul}, in which the authors argue that ground states
with zero and maximum spin will dominate. The evidence from our work is
that this effect does not occur in spin clusters.

We gratefully acknowledge discussions and correspondence with
C. Hamer, A. Sandvik, and V. Zelevinsky and support from the Australian 
Research Council.

\begin{figure}[h]
\vspace{0pt}
\hspace{-35pt}
\epsfxsize=9.cm
\centering\leavevmode\epsfbox{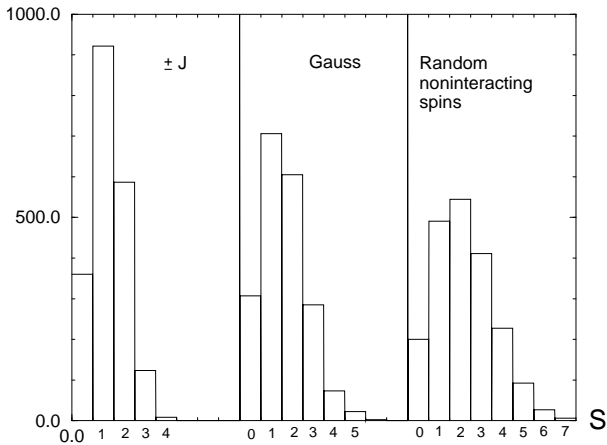}
\vspace{-20pt}
\caption{\it {The ground state spin distributions (N=18, 2000 samples)
for both the $\pm J$ and Gaussian
models, together with the distribution which would be expected in the
absence of interactions.}}
\label{Fig1}
\end{figure}

\begin{figure}[h]
\vspace{0pt}
\hspace{-35pt}
\epsfxsize=9.cm
\centering\leavevmode\epsfbox{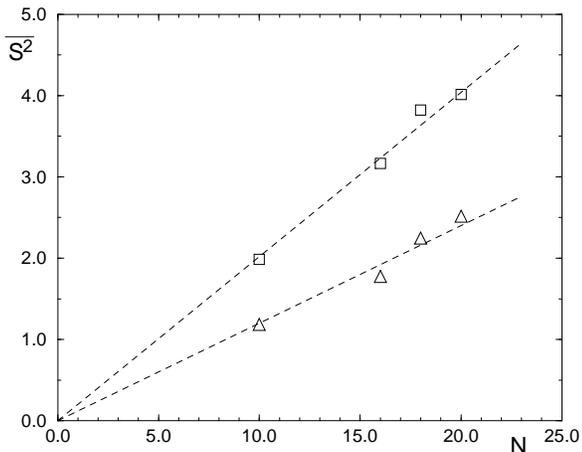}
\vspace{-20pt}
\caption{\it {The average ground state spin squared  versus size of the 
cluster. The triangles correspond to the $\pm J$ model, and the squares
correspond to the Gaussian model.}}
\label{Fig2}
\end{figure}
\noindent

\begin{figure}[h]
\vspace{-20pt}
\hspace{-35pt}
\epsfxsize=9.cm
\centering\leavevmode\epsfbox{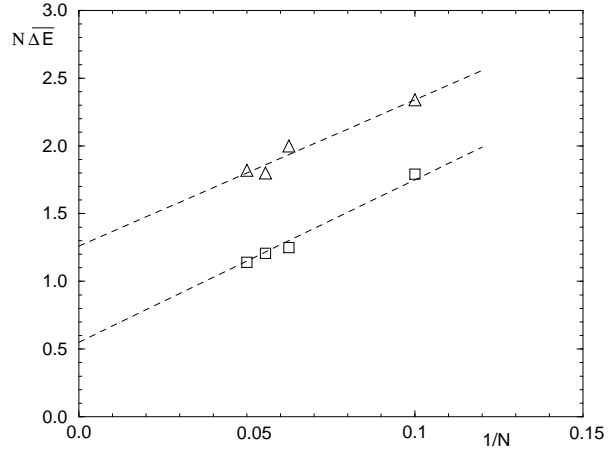}
\vspace{-20pt}
\caption{\it {The average gap ${\overline {\Delta E}}$ multiplied by the size
of the cluster  versus inverse size of the cluster.
The triangles correspond to the $\pm J$ model, and the squares
correspond to the Gaussian model.}}
\label{Fig3}
\end{figure}
\noindent

\vspace{1.cm}

Table I.\\
\vspace{10pt}
\begin{tabular}{|c|c|c|c|c|c|}\hline
 &  N      & 10       & 16      & 18     & 20   \\ \hline
$\pm J$ & $E_0/N$  &  -0.493  & -0.498  & -0.500 & -0.500 \\
model& $m_{sg}^2$ &  0.051  & 0.0355  & 0.0325 & 0.0303 \\
&$f_{+}$&  0.552  & 0.528  & 0.504 & 0.564 \\
&$f_{-}$ &  0.422  & 0.454  & 0.464 & 0.418 \\
&${\overline {\Delta E}}$&  0.234  & 0.125  & 0.100 & 0.091 \\ \hline
Gaussian  & $E_0/N$ &  -0.467  & -0.471  & -0.472 & -0.473 \\
model&$m_{sg}^2$  &  0.050  & 0.0358  & 0.0332 & 0.0306 \\
&$f_{+}$ &  0.512  & 0.484  & 0.466 & 0.510 \\
&$f_{-}$ &  0.474  & 0.504  & 0.512 & 0.462 \\
&${\overline {\Delta E}}$ &  0.179  & 0.078  & 0.067 & 0.057 \\ \hline
\end{tabular}\\
{\it The data  for different cluster sizes N and for two types of 
random distribution. $E_0/N$ is the average ground state energy per 
site; $m^2_{sg}$ is the spin glass order parameter; $f_{+1}$ is the 
fraction of configurations which have spin of the first excitation $S_0+1$, 
where $S_0$ is spin of the ground state; $f_{-1}$ is the fraction of 
configurations  which have spin of the first excitation $S_0-1$;
${\overline {\Delta E}}$ is the average energy gap.}

\vspace{1.cm}

Table II.\\
\vspace{10pt}
\begin{tabular}{|c|c|c|c|c|c|}\hline
$L_x \times L_y$ &$4\times 3$ & $6\times 3$ & $ 3\times 4$ &$4\times 4$ &
$5\times 4$ \\ \hline
 $\pm J$&&&&&\\  
model      & 0.220   &  0.208  & 0.214   & 0.200 & 0.198\\ \hline
Gaussian&&&&&\\
 model   & 0.194   &  0.190  & 0.190   & 0.189 & 0.190\\
 \hline
\end{tabular}\\
{\it The average spin stiffness $\rho_s$ for different rectangular clusters.
The twist is imposed along the x direction.}

\end{document}